\documentclass[%
 reprint,
superscriptaddress,
showkeys,
 amsmath,amssymb,
 aps,
pre, 
longbibliography,
onecolumn
]{revtex4-2}

\usepackage{graphicx}
\usepackage[caption=false]{subfig} 

\usepackage{dcolumn}
\usepackage{bm}

\usepackage[a4paper, total={6in, 9in}]{geometry}

\newcommand{\ee}{\end{equation}} 
\newcommand{\be}{\begin{equation}}

\usepackage[mathscr]{euscript}  
\usepackage{xcolor}  

\usepackage{tcolorbox}
\usepackage{comment}
\usepackage{float}

\makeatletter
\newsavebox{\@brx}
\newcommand{\llangle}[1][]{\savebox{\@brx}{\(\m@th{#1\langle}\)}%
  \mathopen{\copy\@brx\kern-0.5\wd\@brx\usebox{\@brx}}}
\newcommand{\rrangle}[1][]{\savebox{\@brx}{\(\m@th{#1\rangle}\)}%
  \mathclose{\copy\@brx\kern-0.5\wd\@brx\usebox{\@brx}}}
\makeatother

\begin{document}

\preprint{ApS/123-QED}

\title{Anisotropic Brownian particles under resetting}

\author{Subhasish Chaki}\thanks{subhasischaki@gmail.com}

\author{Kristian St\o{}levik Olsen}\thanks{kristian.olsen@hhu.de}

\author{Hartmut L\"{o}wen}

\affiliation{Institut für Theoretische Physik II - Weiche Materie, Heinrich-Heine-Universität Düsseldorf, D-40225 Düsseldorf, Germany}

\begin{abstract} %
\noindent We study analytically the dynamics of an anisotropic particle subjected to different stochastic resetting schemes in two dimensions. The Brownian motion of shape-asymmetric particles in two dimensions results in anisotropic diffusion at short times, while the late-time transport is isotropic due to rotational diffusion. We show that the presence of orientational resetting promotes the anisotropy to late times.  When the spatial and orientational degrees of freedom are reset, we find that a non-trivial spatial probability distribution emerges in the steady state that is determined by the initial orientation, particle asymmetry and the resetting rate. When only spatial degrees of freedom are reset while the orientational degree of freedom is allowed to evolve freely, the steady state is independent of the particle asymmetry. When only particle orientation is reset, the late-time probability density is given by a Gaussian with an effective diffusion tensor, including off-diagonal terms, determined by the resetting rate. 
Generally, the coupling between the translational and rotational degrees of freedom, when combined with stochastic resetting, gives rise to unique behaviour at late times not present in the case of symmetric particles. Considering recent developments in experimental implementations of resetting, our results can be useful for the control of asymmetric colloids, for example in self-assembly processes.

\end{abstract}

\pacs{Valid pACS appear here} 
\maketitle


\section{Introduction}

\noindent A wide variety of processes in physics, chemistry, and biology are comprised of diffusing particles that are highly anisotropic in shape. Prominent examples include rod-like bacteria or viruses \cite{bawden1936liquid,wen1989observation,graf1999phase,doostmohammadi2016defect}, nematic macromolecules \cite{gray1984theory,narayan2007long}, and shape-asymmetric colloids \cite{ivlev2012complex,lowen1994brownian,bolhuis1997tracing,dijkstra2014entropy,lowen1999anisotropic,roller2021observation}. There are several recent research avenues in soft condensed matter, where asymmetric colloidal micro- and nano-particles play a prominent role. In colloidal self-assembly, control of shape-asymmetric particles is crucial for the assembly of materials with novel functions and properties \cite{wu2018anisotropic,glotzer2007anisotropy,sacanna2011shape,thorkelsson2015self}. In nanomedicine, how to externally control the dynamics of an asymmetric colloid moving inside the human body is a central question \cite{simeonidis2016situ}.

\noindent The diffusive motion of shape-asymmetric objects is a topic with a long history, dating back to the work of Perrin almost a century ago \cite{perrin1934mouvement,perrin1936mouvement,koenig1975brownian}. The translational motion of these anisotropic particles can be quite different from those of spherical particles. When left to diffusive, the anisotropic particles show a crossover from short-time anisotropic diffusion, due to the coupling of rotational and translational motion, to effective isotropic diffusion at late times \cite{han2006brownian}. The crossover between these regimes depends on the timescale set by the rotational diffusion coefficient, which determines the directional memory along the particle’s long axis.  
Furthermore, the transient behaviour of an anisotropic particle is longer-lived in two dimensions when compared to the three-dimensional case. This suggests that the transient behaviour may modulate the reaction rate of diffusion-limited reactions in intracellular membranes, where reaction time is frequently of the order of microseconds and the environment is dimensionally restricted \cite{bicout1996stochastic,schnell2004reaction}. Experimental methods are also available for synthesizing and characterizing the dynamics of ellipsoidal colloids \cite{han2006brownian}. Experimentally, it has been shown that the transition from anisotropic to isotropic diffusion in two dimensions occurs over a period of a few seconds for a free micrometer-sized ellipsoidal particle \cite{han2009quasi}.

The Brownian motion of an anisotropic particle has been studied in many situations over the years, including dynamics in confinement and external potentials \cite{han2009quasi,grima2007brownian}, first passage analysis \cite{levernier2016mean, ghosh2022persistence}, active matter \cite{ghosh2020persistence,anchutkin2024run,shemi2018self} and stochastic thermodynamics \cite{marino2016entropy}. So far, to the best of our knowledge, there has been little or no work on anisotropic particles subjected to \emph{stochastic resetting}. Stochastic resetting is a process whereby a system’s state is brought back to its initial condition at a constant rate \cite{evans2011diffusion,evans2011optimal,evans2020review}. A wide range of intriguing phenomena results from this simple rule of intermittent interruption, for example, anomalous relaxation properties \cite{majumdar2015dynamical,gupta2019stochastic} and non-trivial steady states resulting from the confining effect of resetting \cite{pal2015diffusion,nagar2016diffusion}. Over recent years, a myriad of aspects has been investigated, including target search processes under resets \cite{reuveni2016optimal,GM20,faisant2021_2d,pal2020search,durang2019first,tucci2022first,friedman2020exp,toledo2022first,ahmad2022first,ahmad2019first,xu2022stochastic,ray2020space,pal2019first,pal2022inspection,mercado2022reducing,Bressloff2020cyto,Bressloff2020intra}, stochastic thermodynamics \cite{fuchs2016stochastic,pal2017integral,pal2021thermodynamic,gupta2020work,gupta2022work,mori2023entropy,olsen2023thermodynamic,pal2023thermodynamic,olsen2024thermodynamic,goerlich2023experimental}, and active matter \cite{kumar2020active,evans2018run,santra2020run,baouche2024active,goswami2021stochastic}. Recently, the effects of resetting in systems with multiple coupled degrees of freedom have been studied. In particular, the case where an observed degree of freedom experiences indirect effects of resets due to a coupling to a resetting variable has been investigated \cite{abdoli2021stochastic,Olsen_2024}.

\noindent In the majority of past studies, resetting is applied to symmetric or point-like particles. Here, we extend past studies by studying the effect of various resetting schemes applied to a shape-asymmetric Brownian particle. In particular, we consider the case of a Brownian anisotropic particle, where a rotational-translational coupling is present which separates the dynamics from the symmetric counterpart. We study the dynamics of a Brownian anisotropic particle under various resetting schemes, all of which may be implemented experimentally. Resetting of the translational degrees of freedom can be achieved by optical tweezers \cite{tal2020experimental,besga2020optimal,besga2021dynamical,goerlich2023experimental}, while resetting of the particle's orientation can be performed using techniques developed for magnetic anisotropic colloid assembly \cite{tierno2014recent}.  We seek to understand the dynamical consequences of the combined effect of particle asymmetry and resetting in two dimensions, which can be characterised by moments and marginal distributions, available, for example, in colloidal experiments using particle tracking.

\noindent This paper is structured as follows. Section \ref{sec:gentheory} introduced the necessary background theory, including the Brownian motion of anisotropic particles in two dimensions, as well as the renewal approach to stochastic resetting when multiple degrees of freedom is present. We then proceed to investigate the dynamical and stationary properties under various resetting schemes, namely complete resetting of all degrees of freedom, translational resetting, and orientational resetting in sections \ref{sec:complete}, \ref{sec:translational} and \ref{sec:orientation} respectively. Section \ref{sec:discussion} offer a concluding discussion.

\section{General theory}\label{sec:gentheory}

\subsection{Diffusion of an anisotropic particle}

\noindent Here we consider the dynamics of an anisotropic free Brownian particle in two dimensions with mobilities $\Gamma_{||}$ and $\Gamma_{\perp}$ along the longer and the shorter axes of the particle, respectively. Additionally, the motion of the particle is subjected to rotational diffusion, characterized by the mobility  $\Gamma_\theta$. The translational and rotational diffusivities are defined through Einstein-Smoluchowski relations $D_{||} = k_BT \Gamma_{||}$ and $D_\perp = k_B T \Gamma_\perp$, and $D_\theta = k_BT\Gamma_\theta$, where $T$ is the temperature of the bath. The particle at any given time $t$ can be described by the position vector $\vec{r}(t )$ of its centre of mass, which can be decomposed as $(\delta  \Tilde{x}, \delta  \Tilde{y})$ in the body frame and $(\delta x, \delta y)$ in the laboratory frame. The angle between the lab and body frames is denoted as $\theta (t)$, which evolves diffusively as 

\begin{equation}
\begin{split}
 \frac{\partial \theta}{\partial t}= \sqrt{2 k_BT\Gamma_{\theta}}\Tilde{\eta}_\theta (t). 
\label{eq:dynamics_body_frame_angle}
\end{split}
\end{equation} 

\noindent The translational motions of the anisotropic particle are decoupled in the body frame, and are described by the Langevin equations,

\begin{equation}
\begin{split}
 \frac{\partial \Tilde{x}}{\partial t}=\sqrt{2 k_BT\Gamma_\parallel} \Tilde{\eta}_x (t); \,\,\,   \frac{\partial \Tilde{y}}{\partial t}=\sqrt{2 k_BT \Gamma_\perp}\Tilde{\eta}_y (t). 
\label{eq:dynamics_body_frame}
\end{split}
\end{equation} 

\noindent The translational and rotational noise terms $\Tilde{\eta}_x (t)$, $\Tilde{\eta}_y (t)$ and $\Tilde{\eta}_\theta (t)$ in equation (\ref{eq:dynamics_body_frame}) are assumed to be Gaussian and characterized by the following mean and variances

\begin{equation}
\begin{split}
&\langle \Tilde{\eta}_i (t) \rangle=0,  \,\,\, \mathrm{and} \,\,\, \langle \Tilde{\eta}_i (t) \Tilde{\eta}_j \left(t^{\prime}\right) \rangle=  \delta_{ij} \delta\left(t - t^{\prime}\right) 
\label{eq:body_frame_noise}
\end{split}
\end{equation} 

\noindent The displacements in the lab and the body frames are related by the following equations, 

\begin{equation}
\begin{split}
\delta x=\cos\theta \delta  \Tilde{x} - \sin \theta \delta  \Tilde{y},  \\
\delta y=\cos\theta \delta  \Tilde{y} + \sin \theta \delta  \Tilde{x}.  \\
\label{eq:coordinate_transformation}
\end{split}
\end{equation} 

\noindent Dividing the equation (\ref{eq:coordinate_transformation}) by $\delta t$, taking the limit $\delta t \rightarrow 0$, and substituting the translational velocities in the body frame from equation (\ref{eq:dynamics_body_frame}), we get the equations of motion in the lab frame as follows

\begin{equation}
\begin{split}
 \frac{\partial {x}}{\partial t}={\xi}_x (t); \,\,\, \frac{\partial {y}}{\partial t}={\xi}_y(t)
\label{eq:dynamics_lab_frame}
\end{split}
\end{equation} 

\noindent The random variables $\xi_i$ have zero mean and the correlations at fixed $\theta(t)$ are given by

\begin{figure}[t!]
    \centering
    \includegraphics[width = 15cm]{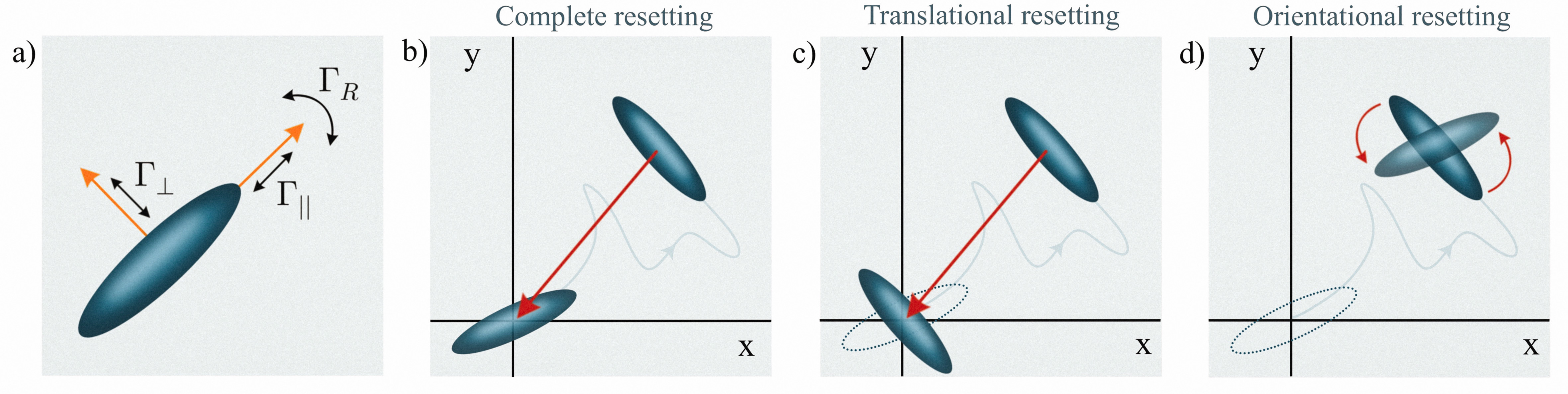}
    \caption{a) Anisotropic particle with three mobilities; parallel to body axis $\Gamma_{||}$, perpendicular to the body axis $\Gamma_\perp$, and a rotational motility $\Gamma_R$. b) Under complete resetting, all degrees of freedom are reset to their initial conditions. c) For translational resets, only the particle's center-of-mass position is reset. d) Under orientational resetting, the particle orientation angle is reset, while the center-of-mass positions remain unchanged. }
    \label{fig:ellipse}
\end{figure}

\begin{equation}
\begin{split}
 \langle {\xi}_i (t) {\xi}_j \left(t^{\prime}\right) \rangle= 2 k_B T \Gamma_{ij} \left[\theta (t)\right] \delta\left(t - t^{\prime}\right) ,
\label{eq:lab_frame_noise}
\end{split}
\end{equation} 

\noindent where $\Gamma_{ij} \left[\theta (t)\right] =\frac{\bar{\Gamma}}{2} + \frac{\Delta \Gamma}{2} \left[\begin{array}{cc} \cos 2\theta(t) & \sin 2\theta(t) \\ \sin 2\theta(t) & -\cos 2\theta (t) \end{array}\right]$  with  $\bar{\Gamma}=\frac{\Gamma_{||} +\Gamma_{\perp}}{2}$   and      $\Delta{\Gamma}=\frac{\Gamma_{||} -\Gamma_{\perp}}{2}$.

\noindent Integrating equation (\ref{eq:dynamics_lab_frame}) with respect to time for $x$ direction, one obtains 

\begin{equation}
\begin{split}
x (t) = \int_0^t dt^\prime \xi_x \left(t^\prime \right) + x_0
\label{eq:lab_frame_x}
\end{split}
\end{equation} 

\noindent The rotational degree of freedom evolves on the unit circle following an ordinary Brownian that is  independent of the translational degrees of freedom. Hence, the second moment of the particle, conditional on an initial orientation angle $\theta_0$, can be explicitly calculated from equation (\ref{eq:lab_frame_x}) as

\begin{equation}
\begin{split}
\langle x^2 (t) | \theta_0\rangle & = \int_0^t dt^\prime \int_0^t dt^{\prime \prime} \langle \xi_i \left(t^\prime \right)  \xi_i \left(t^{\prime\prime} \right) \rangle  \\
&= 2 k_B T  \int_0^t dt^\prime \int_0^t dt^{\prime \prime} \langle \Gamma_{ij} \left[\theta (t^\prime)\right] \rangle_{\eta_\theta} \delta\left(t^\prime - t^{\prime\prime}\right)  \\
&= 2 k_B T  \int_0^t dt^\prime \left[\bar{\Gamma} + \frac{\Delta{\Gamma}}{2} \langle \cos \theta \left(t^\prime\right) |\theta_0\rangle \right] \\
&=  2 k_B T \left[\bar{\Gamma} t + \frac{\Delta{\Gamma} }{2} \cos 2\theta_0  \left(\frac{1-e^{-4D_\theta t}}{4 D_\theta}\right) \right]
\label{eq:2nd_moment_x}
\end{split}
\end{equation}

\noindent Here, and throughout, we assume the particle is initially at the origin $x_0 = y_0 = 0$.  Above, we used the following identity

\begin{equation}
\begin{split}
\langle \cos2 \theta \left(t\right) |\theta_0\rangle = Re\langle e^{i 2 \theta} |\theta_0 \rangle = Re \left[\int d\theta e^{i 2 \theta} \frac{e^{-\frac{\left(\theta - \theta_0\right)^2}{4 D_\theta t}}}{\sqrt{4 \pi D_\theta t}} \right] = \cos( 2\theta_0) \: e^{-4 D_\theta t}
\label{eq:lab_frame_theta} 
\end{split}
\end{equation}

\noindent Similarly, one can explicitly calculate the second moment for $y(t)$ direction

\begin{equation}
\begin{split}
\langle y^2 (t) | \theta_0 \rangle = 2 k_B T \left[\bar{\Gamma} t - \frac{\Delta{\Gamma} }{2} \cos 2\theta_0  \left(\frac{1-e^{-4D_\theta t}}{4 D_\theta}\right) \right]
\label{eq:2nd_moment_y}
\end{split}
\end{equation}

\noindent The first term in equations (\ref{eq:2nd_moment_x}) and (\ref{eq:2nd_moment_y}) is attributed to the diffusion of the center-of-mass and the second term results from the displacement of the particle due to the orientational fluctuations. At late times much larger than the rotational timescale $t\gg D_\theta^{-1}$, the diffusion is isotropic with $\langle x^2(t)\rangle = \langle y^2(t)\rangle = 2 k_B T \bar{\Gamma}t$. At short timescales, however, we have the first-order expansions
\begin{align}
    \langle x^2(t)|\theta_0\rangle  &= 2 k_BT \left(\bar{\Gamma} + \frac{\Delta \Gamma}{2} \cos(2\theta_0) \right) t +\mathcal{O}(t^2)\\
    \langle y^2(t)|\theta_0\rangle  &= 2 k_BT \left(\bar{\Gamma} - \frac{\Delta \Gamma}{2} \cos(2\theta_0) \right) t +\mathcal{O}(t^2)
    \label{eq:moment_short_time}
\end{align}
where we see that the shape asymmetry gives rise to anisotropic diffusion at early times, depending on the particle's initial orientation $\theta_0$. It is interesting to compare the mean squared displacements along the $x$ and $y$ directions. The ratio of the variances is given by $K =\frac{{\langle x^2 (t) |\theta_0 \rangle}}{{\langle y^2 (t) |\theta_0 \rangle}}=\frac{ \left(\bar{\Gamma} + \frac{\Delta \Gamma}{2} \cos(2\theta_0) \right) }{\left(\bar{\Gamma} - \frac{\Delta \Gamma}{2} \cos(2\theta_0) \right) }$. 

\noindent The anisotropy in diffusion also generates non-zero cross-correlation between $x$ and $y$ through their mutual coupling to the orientation, 

\begin{equation}
\begin{split}
\langle x(t) y (t) |\theta_0\rangle = 2 k_B T \left[ \frac{\Delta{\Gamma} }{2} \sin 2\theta_0  \left(\frac{1-e^{-4D_\theta t}}{4 D_\theta}\right) \right]
\label{eq:2nd_moment_xy}
\end{split}
\end{equation}

\noindent It is evident that the correlation between $x$ and $y$ in equation (\ref{eq:2nd_moment_xy}) becomes constant at $t\rightarrow \infty$ due to the decorrelation between $\theta_0$ and $\theta(t)$ after a timescale proportional to $D_\theta^{-1}$.

\subsection{Renewal approach to resetting}\label{sec:renew}
\noindent The anisotropic particle in two dimensions can be described by its center-of-mass positions, $(x,y)$ as well as an orientation angle $\theta$. We collect these three degrees of freedom into a single phase space variable $X \equiv (x,y,\theta)$. Upon stochastic resetting, we must prescribe a rule for how all phase space variables are reset. In general, we assume that upon a reset event the phase space point $X(t)$ resets to
\begin{equation}
    X(t) \longrightarrow X_R(X_0,X(t))
\end{equation}
where the resetting location $X_R(X_0,X(t))$ can be a mix of the coordinates $X(t)$ before the reset and the initial coordinates $X_0$ depending on the type of resetting considered.  For Poissonian resets, i.e. the waiting time between resets follows an exponential distribution with rate $r$, the full phase space propagator $P_r(X,t|X_0)$ can be expressed in terms of a renewal equation
\begin{align}\label{eq:renewal_equation_start}
    P_r(X,t|X0) &= e^{- r t}   P_0(X,t|X_0)  + r \int_0^t d\tau e^{-r\tau} \int d X'  P_r(X',t-\tau|X_0)  P_0(X,\tau|X_R(X_0,X'))
\end{align}
Here, $P_0(X,t|X_0)$ is the propagator of the underlying system in the absence of resetting. The first term in equation (\ref{eq:renewal_equation_start}) corresponds to trajectories where no resets have occurred up to time $t$, which happens with probability $e^{-rt}$. The second term takes into account trajectories where resetting takes place, and decomposed the path at the time of the last reset. By changing the functional form of the resetting point in phase space $X_R(X_0, X(t))$, we can study different resetting schemes and how they affect the dynamics. We consider the following resetting schemes:

\begin{itemize}
    \item[i)] \textbf{Complete resetting:} all phase space variables are reset to their initial values $X_R(X_0,X'(t)) = X_0$.
    \item[ii)] \textbf{Translational resetting:} Only the spatial coordinates are reset, i.e. $X_R(X_0,X'(t)) = (x_0,y_0,\theta'(t))$.
    \item[iii)] \textbf{Orientational resetting:} Only the orientation is reset $X_R(X_0,X'(t)) = (x'(t),y'(t),\theta_0)$.
\end{itemize}

\noindent In the following, we consider these cases separately. In particular, we are interested in the reduced dynamics of one of the spatial coordinates, for which reduced renewal equations are derived in each case.

\section{Complete resetting}\label{sec:complete}

\noindent For complete resetting, the propagator $P_r(X,t|X_0)$ in the presence of Poissonian resetting can be written through the last renewal equation 

\begin{equation}
\begin{split}
P_r(X,t | X_0) = e^{-rt} P_0(X,t|X_0) + r \int_0^t d\tau e^{-r\tau} \int dX_0 P(X_0) P_0(X,\tau|X_0)
\label{eq:renewal_equation}
\end{split}
\end{equation}

\noindent Here we have assumed that the initial condition $X_0$ is drawn from a distribution $P(X_0)$.
The last renewal equation can predict the steady state as

\begin{equation}
\begin{split}
P_{\textrm{st}} (X,t | X_0) = r \int_0^{\infty} d\tau e^{-r\tau} \int dX_0 P(X_0) P(X,\tau|X_0) = r  \int dX_0 P(X_0) \tilde{P_0}(X,r|X_0)
\label{eq:renewal_steady_state}
\end{split}
\end{equation} 

\noindent Here $\tilde{P}(X,r|X_0)$ is the Laplace transform of $P(X,\tau|X_0)$. From the last renewal equation, Eq.(\ref{eq:renewal_equation}), we know that the moments also satisfy a renewal equation 

\begin{equation}
\begin{split}
{\langle X^n (t) \rangle}_r = e^{-rt} {\langle X^n (t) \rangle}_0 + r \int_0^t d\tau e^{-r\tau} \int dX_0 P(X_0) {\langle X^n (\tau) \rangle}_0
\label{eq:renewal_moments}
\end{split}
\end{equation} 

\noindent  For our case, $X$ represents three dynamical variables $\left(x,y, \theta \right)$. For a fixed initial state $X_0 = \left(0,0,\theta_0\right)$, we have

\begin{equation}
\begin{split}
P_r(x,y,\theta,t | 0,0,\theta_0) = e^{-rt} P_0(x,y,\theta,t | 0,0,\theta_0) + r \int_0^t d\tau e^{-r\tau}  P_0(x,y,\theta,\tau | 0,0,\theta_0)
\label{eq:renewal_equation_all}
\end{split}
\end{equation} 

\noindent The effect of resetting on the variable $x$ can be obtained by integrating over $y$ and $\theta$ in the above renewal equation:

\begin{equation}
\begin{split}
P_r(x,t | 0,\theta_0) = e^{-rt} P_0(x,t | 0,\theta_0) + r \int_0^t d\tau e^{-r\tau}  P_0(x, \tau | 0,\theta_0)
\label{eq:renewal_equation_x}
\end{split}
\end{equation} 

\noindent The corresponding moment for $x$ in that case would be 

\begin{equation}
\begin{split}
{\langle x^n (t) |\theta_0 \rangle}_r = e^{-rt} {\langle x^n (t) |\theta_0 \rangle}_0 + r \int_0^t d\tau e^{-r\tau}  {\langle x^n (\tau) \rangle}_{\theta_0}
\label{eq:renewal_moments_x_all}
\end{split}
\end{equation}

\noindent Next, we calculate the second moment of $x$ using equation (\ref{eq:2nd_moment_x}) and $n=2$ in the equation (\ref{eq:renewal_moments_x_all}),

\begin{equation}
\begin{split}
{\langle x^2 (t) |\theta_0 \rangle}_r = \frac{2k_B T \bar{\Gamma} }{r} \left(1-e^{-rt}\right) + \frac{k_B T \Delta \Gamma \cos 2\theta_0}{\left(r+4D_\theta\right)} \left(1-e^{-\left(r+4D_\theta\right) t}\right)
\label{eq:renewal_moments_x_all_ellipse}
\end{split}
\end{equation}  

\noindent Similarly, one can also calculate ${\langle y^2 (t) |\theta_0 \rangle}_r$ and the cross-correlation ${\langle x (t) y(t) |\theta_0 \rangle}_r$ in the presence of resetting. We find

\begin{equation}
\begin{split}
{\langle y^2 (t) |\theta_0 \rangle}_r = \frac{2k_B T \bar{\Gamma} }{r} \left(1-e^{-rt}\right) - \frac{k_B T \Delta \Gamma \cos 2\theta_0}{\left(r+4D_\theta\right)} \left(1-e^{-\left(r+4D_\theta\right) t}\right)
\label{eq:renewal_moments_y_all_ellipse}
\end{split}
\end{equation}

\begin{equation}
\begin{split}
{\langle x(t) y(t) |\theta_0 \rangle}_r = \frac{k_B T \Delta \Gamma \sin 2\theta_0}{\left(r+4D_\theta\right)} \left(1-e^{-\left(r+4D_\theta\right) t}\right)
\label{eq:renewal_moments_xy_all_ellipse}
\end{split}
\end{equation}

\begin{figure}
    \centering
    \includegraphics[width = 14cm]{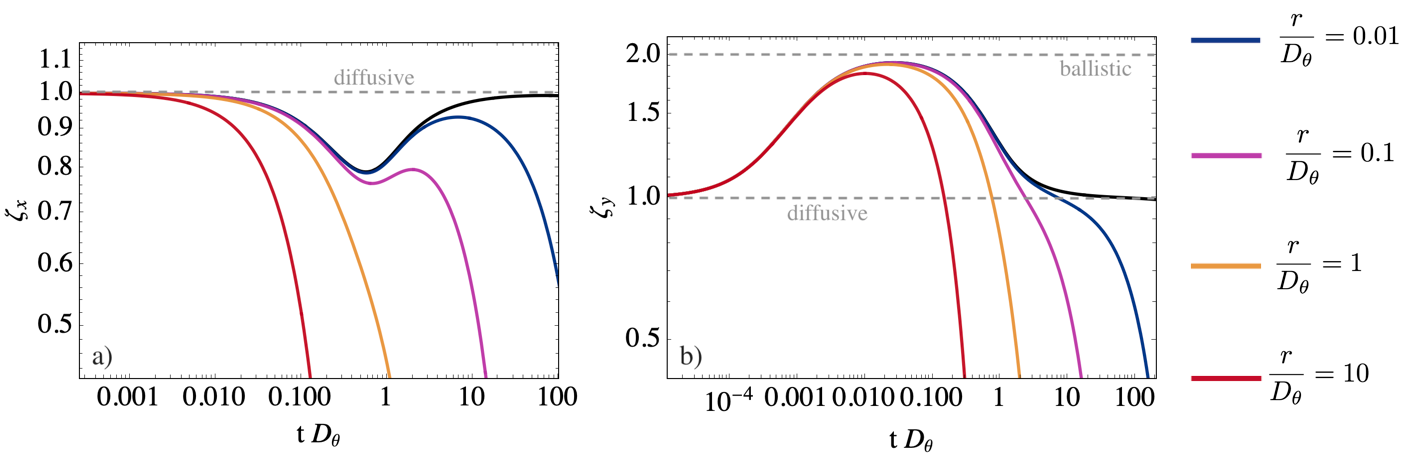}
    \caption{Dynamical exponents $\zeta_x(t)$ and $\zeta_y(t)$ for the mean squared displacement in the $x$ and $y$ direction for various values of resetting rate. The black line shows the exponents without resetting. Parameters are set to $\bar{\Gamma} = 1.0$, $\Delta \Gamma = 0.001$, with initial orientation $\theta_0 = 0$.}
    \label{fig:exponents_all}
\end{figure}

\noindent To gain further insights into the dynamics, we consider the dynamical exponents $\zeta_x(t)=\frac{d \left[\log \langle x^2(t)\rangle\right]}{d \left[ \log t\right]}$ and $\zeta_y(t)=\frac{d \left[\log \langle y^2(t)\rangle\right]}{d \left[ \log t\right]}$ for both $x$ and $y$ directions respectively. These are plotted in Fig. (\ref{fig:exponents_all}). In Fig. (\ref{fig:exponents_all}), we chose the particle initially oriented along the $x$ direction. Without resetting, the exponent $\zeta(t)$ clearly exhibits three distinct regimes \textendash \, short time diffusive regime $\left(\zeta_{x,y}(t) \approx 1\right)$ for both the $x$ and $y$ directions, intermediate superdiffusion $\left(\zeta_y(t) > 1\right)$ for $y$ direction and subdiffusion $\left(\zeta_x(t) < 1\right)$ for $x$ direction and long time diffusive motion $\left(\zeta_{x,y}(t) \approx 1\right)$ for both the $x$ and $y$ directions. A transition to confined motion $\left(\zeta_{x,y}(t) \approx 0\right)$ occurs in the presence of resetting. At a very short times, $t \ll \text{min}(r^{-1}, D_\theta^{-1})$, the particle moves diffusively in each direction with different mobilities $\Gamma_{||}$ and $\Gamma_\perp$ for $\theta_0=0$ as shown in equations (\ref{eq:moment_short_time}), and the motion is not yet affected by resetting. After the characteristic timescale set by rotational diffusion, the particle performs diffusive motion with an effective diffusion constant $\bar{D}$ due to the decorrelation in the orientational degrees of freedom. Since the mobility in the $x$ and $y$ directions are larger and smaller than $\bar{D}$ respectively, the motion in these directions must be decelerated or accelerated to reach the isotropic diffusivity $\bar{D}$ in the long time limit. In the presence of resetting, a transition to confinement ($\zeta_{x,y} \approx 0$) occurs at a typical time $r^{-1}$, potentially interrupting the other dynamical regimes. 

\begin{figure}
    \centering
    \includegraphics[width = 8cm]{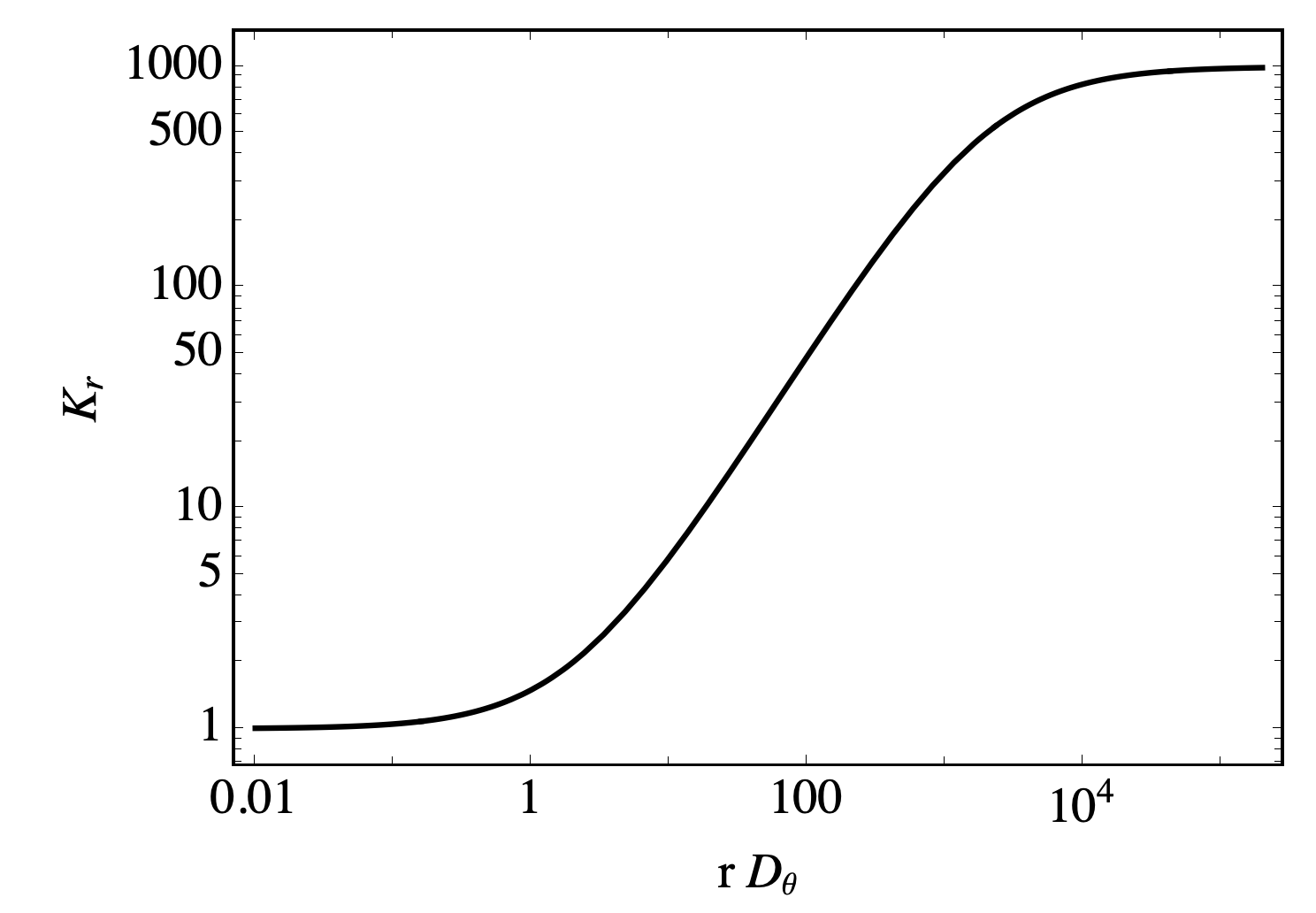}
    \caption{Ratio of variances $K_r$ as a function of resetting rate. The anisotropy grows with resetting rate, and saturates in the $r\to \infty$ limit. }
    \label{fig:Kr}
\end{figure}

Since resetting bring the particle orientation back to its initial angle, the anisotropy found in the short-time expansions Eq. (\ref{eq:moment_short_time}) is partially promoted to late times. To see this, we consider as a measure for the late-time anisotropy the ratio of the mean squared displacements in the $x$ and $y$ directions. In the absence of resetting, the steady-state second moment in the presence of resetting for a spherical Brownian particle with diffusivity $\bar{D}$ is $\frac{2\bar{D}}{r}$ along both the $x$ and $y$ directions. However, the steady-state second moments in the presence of resetting for the anisotropic particle are modified and become larger or smaller than $\frac{2\bar{D}}{r}$ in the $x$ and $y$ directions. Indeed, at late times $ \langle x^2 |\theta_0\rangle = \frac{2\bar{D}}{r}+\frac{k_B T \Delta \Gamma \cos 2\theta_0}{\left(r+4D_\theta\right)}$ and $\langle y^2 |\theta_0\rangle = \frac{2\bar{D}}{r}-\frac{k_B T \Delta \Gamma \cos 2\theta_0}{\left(r+4D_\theta\right)}$. The ratio of the variances is given by
\begin{equation}
    K_r =\frac{{\langle x^2 (t) |\theta_0 \rangle}_r}{{\langle y^2 (t) |\theta_0 \rangle}_r}=\frac{\frac{2\bar{D}}{r}+\frac{k_B T \Delta \Gamma \cos 2\theta_0}{\left(r+4D_\theta\right)}}{\frac{2\bar{D}}{r}-\frac{k_B T \Delta \Gamma \cos 2\theta_0}{\left(r+4D_\theta\right)}}
\end{equation}
In the case of a spherical particle $(\Delta \Gamma = 0)$, one obtains $K_r=1$ independent of the stochastic resetting. The motion is not isotropic for $\Delta \Gamma \neq 0$ under the resetting of all the variables, and the degree of anisotropy increases with $r$ and eventually saturates, as shown in Fig. (\ref{fig:Kr}). The timescale of the cross-correlation between $x$ and $y$ directions is renormalized in the presence of resetting, but the form remains invariant (see Eq.(\ref{eq:2nd_moment_xy}) and Eq. (\ref{eq:renewal_moments_xy_all_ellipse})).

\subsection{Perturbative steady state for near-symmetric particles}

\noindent While we can derive exact expressions for the lower-order moments as discussed above, we would also like to gain insights into the overall shape of the non-equilibrium steady state under resetting. To achieve this, we consider for simplicity a perturbative approach where we calculate the propagator in the absence of resetting, and then use the renewal approach to resetting in order to find the steady state.

\noindent In the absence of resetting, the dynamics can be described by the Smoluchowski-Perrin equation
\begin{equation}
  \partial_t P_0 (X,t|X_0) = D_\theta \partial_\theta^2 P_0 (X,t|X_0)+ \nabla \cdot \left[ \mathbb{D}\cdot \nabla \right] P_0 (X,t|X_0),
\end{equation}

\noindent where $\mathbb{D}$ is the diffusion tensor $\mathbb{D}_{ij} = k_BT \Gamma_{ij} = D_{||} n_in_j + D_\perp (\delta_{ij} - n_in_j)$with $\bm{n} = (\cos \theta, \sin \theta)$ the unit vector pointing along the particle's major axis. Performing a Fourier transform,
\begin{equation}
    \hat P_0 (\bm{k},\theta,t|\theta_0) \equiv \int d\bm{x} e^{- i \bm{k}\cdot \bm{x}} P_0 (\bm{x},\theta,t|\theta_0) ,
\end{equation}
we find the transformed equation

\begin{equation}
  \partial_t \hat P_0 (\bm{k},\theta,t|\theta_0) = D_\theta \partial_\theta^2 \hat P_0(\bm{k},\theta,t|\theta_0) + \bm{k} \cdot \left[ \mathbb{D}\cdot \bm{k} \right] \hat P_0(\bm{k},\theta,t|\theta_0).
\end{equation}

\noindent For simplicity, we here consider the steady state in the $x$-direction. Hence, we consider a wave vector $\bm{k} = k \bm{e}_x$, with $\bm{e}_x$ the unit vector in the $x$-direction. The marginal distribution in the $x$ direction satisfies \cite{mayer2021two}
\begin{equation}
    \partial_t P_0(k,\theta,t|\theta_0) = \left[ D_r \partial_\theta^2 - k^2 \left(\bar{D} +\frac{\Delta D}{2}\cos(2\theta)\right)\right] P_0(k,\theta,t|\theta_0)
\end{equation}

\noindent A natural interpretation of this equation is that the effectively experienced diffusivity in the $x$-direction $\bar{D} +\frac{\Delta D}{2}\cos(2\theta)$ depends on the orientation of the particle. To find an analytical solution that provides some insight into the shape of the steady state under resetting, we consider a perturbative approach, where we expand in power of the dimensionless asymmetry parameter $\varepsilon \equiv \Delta D /\bar{D}$. We write the marginal Smoluchowski-Perrin equation as

\begin{equation}
    \partial_t P_0(k,\theta,t|\theta_0) =  D_r \partial_\theta^2\hat P(k,\theta,t|\theta_0) - k^2 \bar{D} \hat P_0 - \varepsilon  \: k^2 \bar{D}  \frac{\cos(2\theta)  }{2}\hat P_0(k,\theta,t|\theta_0)
\end{equation}

\noindent We expand the solution in powers of the asymmetry as

\begin{equation}
    P_0(k,\theta,t|\theta_0) = \sum_{n=0}^\infty p_n(k,\theta,t|\theta_0) \varepsilon^n.
\end{equation}

\noindent where $p_n(k,\theta,t|\theta_0)$ is the coefficient function of the power series.  Inserting this ansatz into the Smoluchowski-Perrin equation gives the coupled equations

\begin{align}
    \partial_t p_0(k,\theta,t|\theta_0) &=  D_r \partial_\theta^2 p_0(k,\theta,t)  - k^2 \bar{D} p_0(k,\theta,t|\theta_0) ,\\
     \partial_t p_n(k,\theta,t|\theta_0) &= -[D_r \partial_\theta^2   + k^2 \bar{D} ] p_n(k,\theta,t|\theta_0) - \frac{k^2 \bar{D} \cos(2\theta)}{2}p_{n-1}(k,\theta,t|\theta_0)
\end{align}

\noindent First, we note that $p_0(k,\theta,t) $ is nothing but the Gaussian solution of a symmetric colloid. As we are interested in the spatial steady state to first order, we integrate over the angular degree of freedom for $n=0,1$ above, resulting in 

\begin{align}
    \partial_t p_0(k,t) &=    - k^2 \bar{D} p_0(k,t) ,\\
     \partial_t p_1(k,t|\theta_0) &= -  k^2 \bar{D}  p_1(k,t|\theta_0) - k^2\bar{D} p_0(k,t)  \int d\theta \frac{\cos(2\theta)}{2} \frac{\exp\left(-\frac{(\theta-\theta_0)^2}{4 D_\theta t}\right)}{\sqrt{4 \pi D_\theta t}}
\end{align}

\noindent where we used the fact that the zeroth order solution is Gaussian. Here the zeroth order solution reads

\begin{equation}\label{eq:zerothorder}
    p_0(k,t) = e^{- \bar{D} k^2 t}
\end{equation}

\noindent The integration in the equation for the first order correction can be carried out exactly, yielding

\begin{equation}
     \partial_t p_1(k,t|\theta_0) = -  k^2 \bar{D}  p_1(k,t|\theta_0) - k^2 \bar{D} \cos(2\theta_0) 
    p_0(k,t) \frac{e^{- 4 D_\theta t}}{2}
\end{equation}

\noindent Using Eq. (\ref{eq:zerothorder}) and performing a Laplace transform results in
\begin{equation}
     \hat{\tilde p}_1(k,s|\theta_0)  = - \frac{k^2 \bar{D} \cos(2\theta_0)}{2(s + 4 D_\theta + \bar{D}k^2)(s + k^2 \bar{D})}
\end{equation}

\noindent Inverting the Fourier transform, we finally arrive at the first order correction to the propagator as
\begin{equation}
    \tilde p_1(x,s|\theta_0) = \cos(2\theta_0)\frac{\alpha(s)}{16 D_\theta } e^{-\alpha(s) |x|} - \cos(2\theta_0)\frac{\alpha(s + 4 D_\theta )}{16 D_\theta  }  e^{-\alpha(s+4D_\theta)|x|}
\end{equation}

\noindent where we introduced the inverse lengthscale $\alpha(s) \equiv \sqrt{s/\bar{D}}$. Since $p_0$ takes care of normalization, $p_1$ integrates to zero as required.

\begin{figure}
    \centering
    \includegraphics[width = 15.5cm]{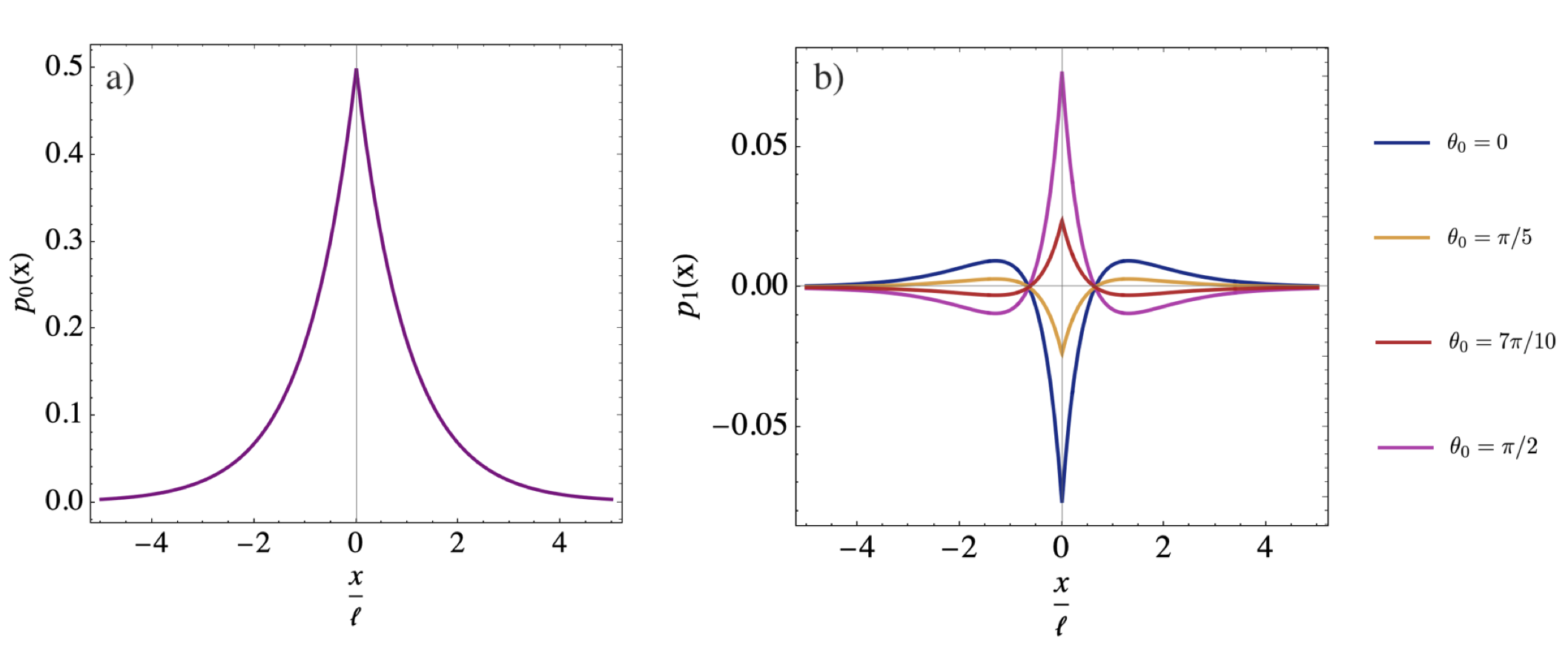}
    \caption{Steady state marginal density under complete resetting. a) Zeroth order solution corresponding to the standard Laplacian solution for a normal Brownian particle with diffusivity $\bar{D}$. b) First order correction to the steady state for a Brownian anisotropic particle for  various $\theta_0$. The lengthscale is set by $\ell^2 = 2 k_B T \bar{\Gamma}/D_\theta$, and the parameters are set to $D_\theta = \bar{D}= r = 1$. }
    \label{fig:perturbative}
\end{figure}

\begin{figure}
    \centering
    \includegraphics[width = 15.5cm]{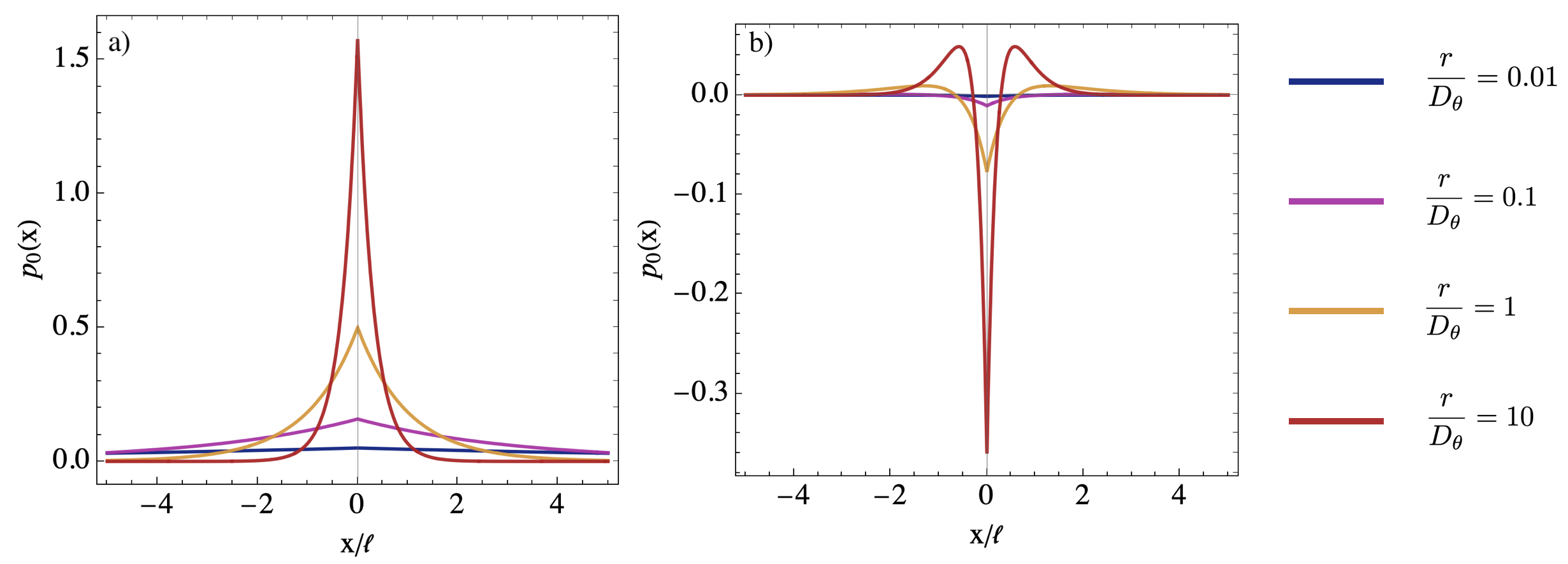}
    \caption{Steady state marginal density under complete resetting. a) Zeroth order solution at different resetting rates. b) First order correction to the steady state for a Brownian anisotropic particle for  various resetting rates.The lengthscale is set by $\ell^2 = 2 k_B T \bar{\Gamma}/D_\theta$, and the parameters are set to $D_\theta = \bar{D}= r = 1$, $\theta_0 = 0$. }
    \label{fig:perturbative2}
\end{figure}

\noindent We can now easily obtain the steady state under the effect of resetting by using the renewal equation Eq. (\ref{eq:renewal_equation_x}), and letting $t \to \infty$. This immediately yields
\begin{equation}
    P_r(x|\theta_0) = \frac{\alpha(r)}{2}e^{-\alpha(r) |x|} + \varepsilon\cos(2\theta_0) \left[  \frac{r \alpha(r)}{16 D_\theta  } e^{-\alpha(r) |x|} - \frac{r \alpha(r + 4 D_\theta )}{16 D_\theta  }  e^{-\alpha(r + 4D_\theta)|x|} \right].
\end{equation}

\noindent Several things are worth noting. The zeroth order solution of the steady state probability distribution is determined by a single length scale $\alpha(r)$ set by resetting. It shows a peak at $x=0$ in Fig. (\ref{fig:perturbative} a) due to the confining effect of the resetting towards the origin. This peak value increases with increasing the resetting rate $r$, as is expected. Once particle asymmetry is introduced, the solution also depends on the length scale set by $\alpha(r + 4 D_\theta)$, and in addition depends on the initial orientation $\theta_0$. Here, the rotational dynamics enter into the spatial steady state due to the coupling between the translational and rotational degrees of freedom, which persists to late times as discussed above.  Fig. (\ref{fig:perturbative} b) shows that the first order correction to the distribution is negative at $x=0$ for $\theta_0 =0$, which is consistent with the enhanced motion along the $x$ direction; hence the steady state widens in this case. When $\theta_0 = \pi/2$, the long axis of the anisotropic particle will be along the $y$ axis, resulting in the suppression of the mobility along the $x$ direction, and consequently, the first order correction becomes positive at $x=0$. The peak value of the first-order correction also increases with the resetting rate $r$, as seen in Fig. (\ref{fig:perturbative2}). This is consistent with the observation that the degree of anisotropy $K_r$ increases with increasing resetting rate $r$.

\section{Translational resetting}\label{sec:translational}

\noindent Next, we consider the case where $x$ and $y$ undergo resets while the orientational degree of freedom $\theta$ variable does not reset. 
Following the discussion in section \ref{sec:renew}, we have the renewal equation

\begin{equation}
\begin{split}
P_r(x,y,\theta,t | 0,0,\theta_0) &= e^{-rt} P_0(x,y,\theta,t | 0,0,\theta_0) \\
&+ r \int_0^t d\tau e^{-r\tau} \int dx^\prime \int dy^\prime \int d\theta^\prime P_r(x^\prime, y^\prime, \theta^\prime,t-\tau|0, 0,\theta_0) P_0(x,y,\theta,\tau | 0,0,\theta^\prime)
\label{eq:renewal_equation_x_and_y}
\end{split}
\end{equation} 

\noindent The propagator for $x$ for this case can be obtained by integrating the above equation over $y$, $\theta$, $x^\prime$ and $y^\prime$, 

\begin{equation}
\begin{split}
P_r(x,t | 0,\theta_0) &= e^{-rt} P_0(x,t | 0,\theta_0) \\
&+ r \int_0^t d\tau e^{-r\tau} \int d\theta^\prime P_0( \theta^\prime,t-\tau|\theta_0) P_0(x,\tau | 0,\theta^\prime)
\label{eq:renewal_equation_x_x_and_y}
\end{split}
\end{equation} 

\noindent The corresponding moment for $x$ in that case would be 

\begin{equation}
\begin{split}
{\langle x^n (t) | \theta_0 \rangle}_r = e^{-rt} {\langle x^n (t) |\theta_0 \rangle}_0 + r \int_0^t d\tau e^{-r\tau}  \int d\theta^\prime P_0( \theta^\prime,t-\tau|\theta_0) {\langle x^n (\tau) | \theta^\prime\rangle}
\label{eq:renewal_moments_x_x_and_y}
\end{split}
\end{equation}

\noindent Next, we calculate the second moment of $x$ for the anisotropic particle using equation (\ref{eq:2nd_moment_x}) and $n=2$ in equation (\ref{eq:renewal_moments_x_x_and_y}), resulting in 

\begin{equation}
\begin{split}
{\langle x^2 (t) |\theta_0 \rangle}_r &= e^{-rt}  2 k_B T \left[\bar{\Gamma} t + \frac{\Delta{\Gamma} }{2} \cos 2\theta_0  \left(\frac{1-e^{-4D_\theta t}}{4 D_\theta}\right) \right] \\
& + r \int_0^t d\tau e^{-r\tau}  \int d\theta^\prime \frac{e^{-\frac{\left(\theta^\prime - \theta_0\right)^2}{4 D_\theta (t-\tau)}}}{\sqrt{4 \pi D_\theta (t-\tau)}}  2 k_B T \left[\bar{\Gamma} \tau + \frac{\Delta{\Gamma} }{2} \cos 2\theta^\prime  \left(\frac{1-e^{-4D_\theta \tau}}{4 D_\theta}\right) \right] \\
&=\frac{2k_B T \bar{\Gamma}}{r} -\left(\frac{2k_B T \bar{\Gamma}}{r} - \frac{k_B T \Delta\Gamma \cos2\theta_0}{4D_\theta-r}\right)  e^{-rt} - \frac{k_B T \Delta\Gamma \cos2\theta_0}{4D_\theta-r} e^{-4D_\theta t}
\label{eq:renewal_moments_x_x_and_y_ellipse}
\end{split}
\end{equation}  

\noindent Similarly, one can also calculate ${\langle y^2 (t) |\theta_0 \rangle}_r$ and the cross-correlation ${\langle x (t) y(t) |\theta_0 \rangle}_r$ in the presence of resetting. We find

\begin{equation}
\begin{split}
{\langle y^2 (t) |\theta_0 \rangle}_r = \frac{2k_B T \bar{\Gamma}}{r} -\left(\frac{2k_B T \bar{\Gamma}}{r} + \frac{k_B T \Delta\Gamma \cos2\theta_0}{4D_\theta-r}\right)  e^{-rt} + \frac{k_B T \Delta\Gamma \cos2\theta_0}{4D_\theta-r} e^{-4D_\theta t}
\label{eq:renewal_moments_y_x_and_y_ellipse}
\end{split}
\end{equation}  

\begin{equation}
\begin{split}
{\langle x (t) y(t) |\theta_0 \rangle}_r &=  \frac{2k_B T \Delta\Gamma \sin2\theta_0}{2(4D_\theta -r)} \left(e^{-rt} -e^{-4D_\theta t} \right)
\label{eq:renewal_moments_xy_x_and_y_ellipse}
\end{split}
\end{equation} 

\noindent Note that in the limit $r \to 4D_\theta$, the above equations
remain well-defined if the limit is taken carefully. In the long time limit,  both ${\langle x^2 (t) |\theta_0 \rangle}_r $  and ${\langle y^2 (t) |\theta_0 \rangle}_r $  approach $\frac{2k_B T \bar{\Gamma}}{r}$, and the steady state ratio of the variances, $K_r$ becomes $1$. The cross-correlation ${\langle x (t) y(t) |\theta_0 \rangle}_r$ will be zero in the long time limit, but at intermediate times, ${\langle x (t) y(t) |\theta_0 \rangle}_r$ shows a non-monotonic behaviour against time. The initial growth is similar to what is seen in the absence of resetting, while resets exponentially decrease the correlations at later times.
Therefore, as the state at late time has no cross-correlations and no anisotropy, the steady-state distribution for $x$ will become $P_r(x)=\frac{\alpha(r)}{2}\exp \left(-\alpha(r)|x|\right)$, with $\alpha(r) = \sqrt{r/\bar{D}}$, which is independent of $D_\theta$. Due to isotropy at late time, the distribution in the $y$ direction will be identical.

\section{Orientational resetting}\label{sec:orientation}

\noindent Finally, we consider the case of orientational resetting, where $\theta$ undergoes resets while the $x$ and $y$ variables do not undergo resets. Again, we follow the general discussion in section \ref{sec:renew}, leading to the last renewal equation 

\begin{equation}
\begin{split}
P_r(x,y,\theta,t | 0,0,\theta_0) &= e^{-rt} P_0(x,y,\theta,t | 0,0,\theta_0) \\
&+ r \int_0^t d\tau e^{-r\tau} \int dx^\prime \int dy^\prime \int d\theta^\prime P_r(x^\prime, y^\prime, \theta^\prime,t-\tau|0, 0,\theta_0) P_0(x,y,\theta,\tau | x^\prime, y^\prime,\theta_0)
\label{eq:renewal_equation_theta}
\end{split}
\end{equation} 

\noindent At the time of the last resetting $t-\tau$, the system is in some arbitrary state $\left(x^\prime, y^\prime, \theta^\prime\right)$. As only $\theta$ is reset to $\theta_0$, the system proceeds to evolve towards $\left(x, y, \theta\right)$ from the new initial condition $\left(x^\prime, y^\prime, \theta_0\right)$. 
The propagator for $x$ for this case can be obtained by integrating the above equation over $y$, $y^\prime$ $\theta$,  and $\theta^\prime$, 

\begin{equation}
\begin{split}
P_r(x,t | 0,\theta_0) &= e^{-rt} P_0(x,t | 0,\theta_0) \\
&+ r \int_0^t d\tau e^{-r\tau} \int dx^\prime P_r(x^\prime,t-\tau|0,\theta_0) P_0(x,\tau | x^\prime, \theta_0)
\label{eq:renewal_equation_theta_x}
\end{split}
\end{equation} 

\noindent In the case of complete and translational resetting, the renewal equations for the spatial moments could be expressed explicitly in terms of the corresponding densities without resetting. However, in the present case the probability density $P_r(x,t | 0,\theta_0)$ is not given explicitly, as it appears both on the left-hand-side and inside the integral on the right-hand-side of the renewal equation.

In order to calculate the late-time moments, one can decouple the equation through the use of Fourier transforms in space and a Laplace transform in time under the assumption of spatial homogeneity. Recently, Ref. \cite{Olsen_2024} studied the late-time behaviour of the moments for this class of problems, where it was shown that the effective diffusion coefficients take the form 
\begin{align}
    D_\text{eff}^{(x)} &= \lim_{t\to \infty}\frac{{\langle x^2(t) \rangle}_r}{2 t} =\frac{r^2}{2} \mathcal{L}_r [\langle x^2(t) |\theta_0\rangle_0]   \\
    D_\text{eff}^{(y)} &= \lim_{t\to \infty}\frac{{\langle y^2(t) \rangle}_r}{2 t} = \frac{r^2}{2} \mathcal{L}_r [\langle y^2(t) |\theta_0\rangle_0] 
\end{align}
where $\mathcal{L}_r$ denotes a Laplace transform evaluated at $r$. Using the mean squared displacement for the problem without resetting, we immediately arrive at 
\begin{align}
    D_\text{eff}^{(x)} &= k_B T\left(\bar{\Gamma}+\frac{r\Delta \Gamma\cos 2\theta_0}{2(r+4D_\theta)}\right) \label{eq:deffx}\\
    D_\text{eff}^{(y)} &= k_B T\left(\bar{\Gamma}-\frac{r\Delta \Gamma\cos 2\theta_0}{2(r+4D_\theta)}\right)\label{eq:deffy}
\end{align}

\noindent In order to obtain a full time-dependent solution of the moments, we make use of the fact that the positions are themselves not coupled, but determined solely by the orientation $\theta$. Hence, the mean squared displacement takes the same mathematical form as in the case of no resetting, Eq. (\ref{eq:2nd_moment_x}), but now with the orientational dynamics subject to resets:

\begin{equation}
\begin{split}
{\langle x^2 (t) |\theta_0 \rangle}_r = 2 k_B T  \int_0^t dt^\prime \left[\bar{\Gamma} + \frac{\Delta{\Gamma}}{2} {\langle \cos \theta \left(t^\prime\right) |\theta_0\rangle}_r \right] 
\label{eq:2nd_moment_x_theta}
\end{split}
\end{equation}

\noindent To calculate ${\langle \cos \theta \left(t^\prime\right)  |\theta_0\rangle}_r$ in the presence of resetting, we use the last renewal equation only for $\theta$, which takes the simple form

\begin{equation}
\begin{split}
P_r(\theta,t | \theta_0) = e^{-rt} P_0(\theta,t|\theta_0) + r \int_0^t d\tau e^{-r\tau}  P_0(\theta,\tau|\theta_0)
\label{eq:renewal_equation_theta}
\end{split}
\end{equation} 

\noindent  Multiplying with $\cos 2\theta$ and integrating results in 

\begin{equation}
\begin{split}
{\langle \cos 2\theta (t)  |\theta_0\rangle}_r &= e^{-rt} {\langle \cos 2\theta (t)  |\theta_0\rangle}_0 + r \int_0^t d\tau e^{-r\tau}  {\langle \cos 2\theta (\tau)  |\theta_0\rangle}_0 \\
&= \cos (2\theta_0)  e^{-\left(r+4D_\theta\right)t} +\frac{r\cos 2\theta_0}{r+4D_\theta} \left(1-e^{-\left(r+4D_\theta\right)t}\right)
\label{eq:renewal_equation_theta_moment_cos}
\end{split}
\end{equation}

\noindent Using Eq. (\ref{eq:2nd_moment_x_theta}), we find

\begin{equation}
\begin{split}
{\langle x^2 (t) |\theta_0 \rangle}_r &=  2 k_B T  \int_0^t dt^\prime \left[\bar{\Gamma} + \frac{\Delta{\Gamma}}{2} {\langle \cos \theta \left(t^\prime\right)  |\theta_0 \rangle}_r \right] \\
&=2k_B T \left[\left(\bar{\Gamma}+\frac{r\Delta \Gamma\cos 2\theta_0}{2(r+4D_\theta)}\right)t + \frac{2D_\theta \Delta\Gamma\cos 2\theta_0}{\left(r+4D_\theta\right)^2}\left(1-e^{-\left(r+4D_\theta\right)t}\right)\right]
\label{eq:renewal_moments_x_theta}
\end{split}
\end{equation}

\begin{figure}
    \centering
    \includegraphics[width = 14cm]{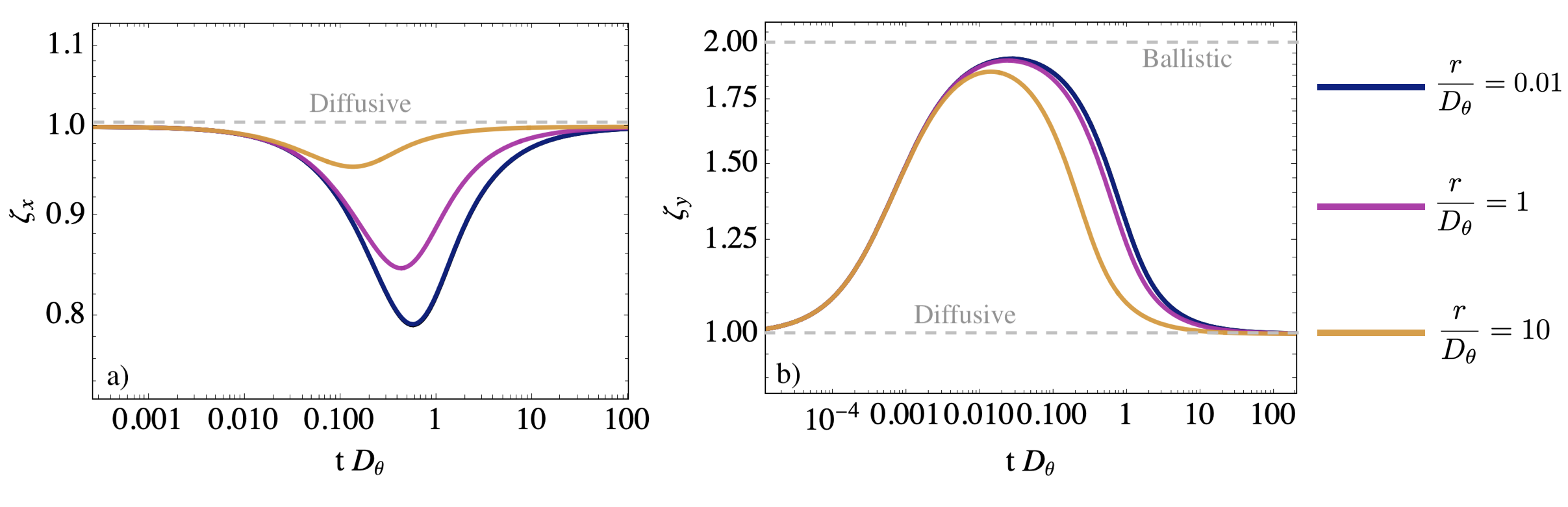}
    \caption{Dynamical exponents $\zeta_x(t)$ and $\zeta_y(t)$ for the mean squared displacement in the $x$ and $y$ direction for various values of resetting rate, in the case of orientational resets. The black line shows the exponents without resetting. Parameters are set to $\bar{\Gamma} = 1.0$, $\Delta \Gamma = 0.001$, with initial orientation $\theta_0 = 0$.}
    \label{fig:exponents_theta}
\end{figure}

\noindent Similarly, one can also calculate ${\langle y^2 (t) |\theta_0 \rangle}_r$ and the cross-correlation ${\langle x (t) y(t) |\theta_0 \rangle}_r$ in the presence of resetting, resulting in 

\begin{equation}
\begin{split}
{\langle y^2 (t) |\theta_0 \rangle}_r =2k_B T \left[\left(\bar{\Gamma}-\frac{r\Delta \Gamma\cos 2\theta_0}{2(r+4D_\theta)}\right)t - \frac{2D_\theta\Delta\Gamma\cos 2\theta_0}{\left(r+4D_\theta\right)^2}\left(1-e^{-\left(r+4D_\theta\right)t}\right)\right]
\label{eq:renewal_moments_y_theta}
\end{split}
\end{equation}  

\begin{equation}
\begin{split}
{\langle x (t) y(t) |\theta_0 \rangle}_r &=  2 k_B T  \int_0^t dt^\prime \left[ \frac{\Delta{\Gamma}}{2} {\langle \sin \theta \left(t^\prime\right) |\theta_0 \rangle}_r \right] \\
&=2k_B T \left[\frac{r\Delta\Gamma\sin 2\theta_0}{2\left(r+4D_\theta\right)}t +  \frac{2D_\theta \Delta\Gamma\sin 2\theta_0}{\left(r+4D_\theta\right)^2}\left(1-e^{-\left(r+4D_\theta\right)t}\right)\right]
\label{eq:renewal_moments_xy_theta}
\end{split}
\end{equation} 

\noindent In the long time limit, since all expressions grow linearly in time, we can neglect the constant offset in the above equations, and the variances of the particle's position in the $xy$ plane and the cross-correlations between $x$ and $y$ can be written as 

\begin{equation}
\begin{split}
{\langle x^2 (t) |\theta_0 \rangle}_r \approx 2k_B T \left(\bar{\Gamma}+\frac{r\Delta \Gamma\cos 2\theta_0}{2(r+4D_\theta)}\right)t
\label{eq:renewal_moments_x_theta_long_time}
\end{split}
\end{equation} 

\begin{equation}
\begin{split}
{\langle y^2 (t) |\theta_0 \rangle}_r \approx 2k_B T \left(\bar{\Gamma}-\frac{r\Delta \Gamma\cos 2\theta_0}{2(r+4D_\theta)}\right)t 
\label{eq:renewal_moments_y_theta_long_time}
\end{split}
\end{equation}  
which agrees with the predictions made in Eqs. (\ref{eq:deffx}) and (\ref{eq:deffy}). The transient behaviour can once again be understood by considering the dynamical exponents of the motion in the $x$ and $y$ directions.  Fig (\ref{fig:exponents_theta}) shows these exponents for the case $\theta_0 = 0$. In this case, the motion in the $x$ direction undergoes a subdiffusive regime before re-entering a diffusing regime at late times. The dynamics in the $y$ direction conversely goes through superdiffusive regime, before also re-entering a diffusive regime.

In contrast to the other resetting schemes considered so far, we now also have a cross correlation that at late times grows as
\begin{equation}
\begin{split}
{\langle x (t) y(t) |\theta_0 \rangle}_r \approx 2k_B T \frac{r\Delta\Gamma\sin 2\theta_0}{2\left(r+4D_\theta\right)}t 
\label{eq:renewal_moments_xy_theta_long_time}
\end{split}
\end{equation} 
Linear growth of the cross correlations has also been seen, for example, in multithermostat Brownian systems under the effect of a Lorentz force \cite{abdoli2020correlations}.

The steady state ratio of the variances, $K_r$, in the presence of resetting is given as 
\begin{equation}
\begin{split}
K_r =\frac{{\langle x^2 (t) |\theta_0 \rangle}_r}{{\langle y^2 (t) |\theta_0 \rangle}_r}=\frac{\left(\bar{\Gamma}+\frac{r\Delta \Gamma\cos 2\theta_0}{2(r+4D_\theta)}\right)}{\left(\bar{\Gamma}-\frac{r\Delta \Gamma\cos 2\theta_0}{2(r+4D_\theta)}\right)}
\label{eq:renewal_moments_xy_theta_ratio}
\end{split}
\end{equation}

\begin{figure}
    \centering
    \includegraphics[width = 15.5 cm]{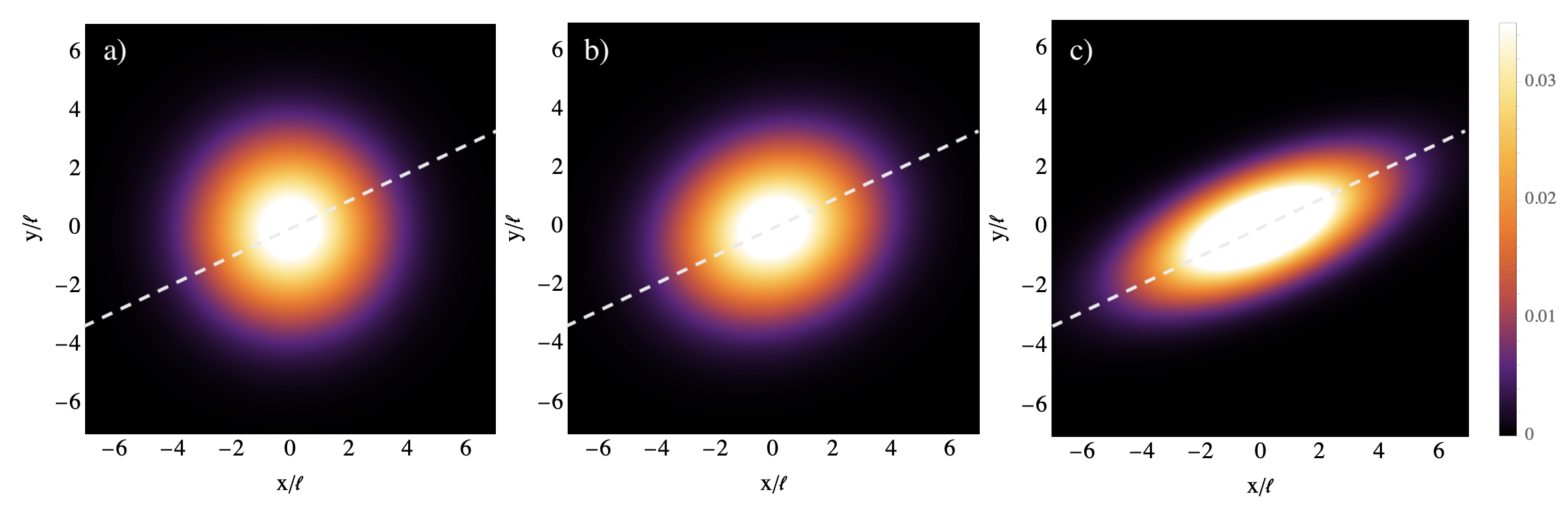}
    \caption{Fixed-time ($ t D_\theta = 4$) two-dimensional positional densities under orientational resetting at different rates: $r = 0$ (panel a), $r = 2 D_\theta$ (panel b) and $r = 5 D_\theta$ (panel c). Lengthscale is set by $\ell^2 = 2 k_B T \bar{\Gamma}/D_\theta$. The dashed line shows initial angle $\theta_0$. Parameters are set to $\theta_0 = \pi/8, D_\theta = 1. \Gamma_{||} = 1$, $\Gamma_\perp = 0.001$.}
    \label{fig:densities}
\end{figure}

\noindent Using $\bar{D} = k_BT \bar{\Gamma}$, we see that this is exactly the degree of anisotropy as in the case of complete resetting. While the degree of anisotropy is the same, the dynamics is different. In particular, we obtain a non-zero cross-correlation between $x$ and $y$ which also grows linearly with time, while for complete resetting the same cross correlations saturates. 

In the long time limit, we can expect that the two-dimensional probability distribution in the Cartesian components will be Gaussian due to the central limit theorem, given as 

\begin{equation}
\begin{split}
P(x,y,t)=\frac{1}{4\pi t \sqrt{\textrm{Det}(\mathbb{D}_\text{eff})}} \exp\left[-\frac{1}{4 t} \left(x,y\right)^{\textrm{T}} \mathbb{D}_\text{eff}^{-1} \left(x,y\right)\right] ,
\label{eq:renewal_moments_xy_theta_long_time_distribution}
\end{split}
\end{equation} 

\noindent where the effective diffusion tensor is given as 
\begin{equation}
    \mathbb{D}_\text{eff} = 2 k_B T \left[\begin{array}{cc} \bar{\Gamma}+\frac{r\Delta \Gamma\cos 2\theta_0}{2(r+4D_\theta)} & \frac{r\Delta\Gamma\sin 2\theta_0}{2\left(r+4D_\theta\right)} \\ \frac{r\Delta\Gamma\sin 2\theta_0}{2\left(r+4D_\theta\right)} & \bar{\Gamma}-\frac{r\Delta \Gamma\cos 2\theta_0}{2(r+4D_\theta)} \end{array}\right].
\end{equation}

\noindent The marginal probability distribution for $x$ can be obtained by integrating the $y$ variable, given as 

\begin{equation}
\begin{split}
P(x,t)=\int dy P(x,y,t) = \frac{1}{\sqrt{4\pi D_\text{eff}^{(x)} t}} \exp\left[-\frac{x^2}{4  D_\text{eff}^{(x)}t} \right] 
\label{eq:renewal_moments_xy_theta_long_time_distribution_x}
\end{split}
\end{equation} 

\noindent  The same argument holds for the marginal probability distribution for $y$. However, the long-time probability distribution in equation (\ref{eq:renewal_moments_xy_theta_long_time_distribution}) cannot be separated into a product of two independent distributions in the $x$ and $y$ directions. Fig. (\ref{fig:densities}) shows the density plots of the steady state probability distribution of the position of the particle. The steady state probability distribution is isotropic without resetting, and the degree of anisotropy grows along the direction of initial angle $\theta_0$ (dashed line in Fig. (\ref{fig:densities}) ).

\noindent It is worth considering a somewhat more intuitive approach to understand the unboundedness of the cross-correlation between $x$ and $y$ with time. We consider the correlation coefficient defined as $C(t)=\frac{\langle x (t) y(t) |\theta_0 \rangle}{\sqrt{\langle x^2 (t) |\theta_0 \rangle}\sqrt{\langle y^2 (t) |\theta_0 \rangle}}$. In the presence of only orientational resetting, $C(t)$ will become constant in the long-time limit, as can be readily verified by using the above results. This is similar to the case of the resetting all the variable, where $C(t)$ will also be a constant, which is an immediate consequence of the fact that a steady state is reached in this case.
This non-zero correlation between the two spatial directions results from the presence of orientational resetting, while for translational resetting, $C(t)$ will be zero in the long-time limit. 

\section{Discussion}\label{sec:discussion}

\noindent In this article, we have explored the effects of shape asymmetry on the two-dimensional diffusive motion of an anisotropic particle in the presence of stochastic resetting. In particular, we studied the short- and long-time behaviour of moments, cross-correlations and steady-state distributions under various resetting protocols. It is well known that anisotropic diffusion persists only for very short times, and that isotropic diffusion is recovered at long-times. Hence, for practical purposes the short-time behaviour can often be neglected and the movement of an asymmetric particle can be described by the Langevin equations for a point particle with an isotropic translational diffusion coefficient given by the average of the diffusion coefficients along the major and minor axes of the particle. This has been shown to be the case for example for a free particle and in a presence of a harmonic confinement in two dimensions \cite{grima2007brownian}. 

We have shown that these assumptions are generally not valid for resetting schemes that include orientational resets, and can be restored only if we apply resetting of the translational degrees of freedom. In this case, the steady-state distribution, moments and cross-correlations are exactly equal to the case of a spherical particle in the presence of stochastic resetting.
On the other hand, when orientational resetting takes place, the short-time anisotropy of the motion is promoted to late times. Both the case of complete resetting and orientational resetting display this, with the degree of anisotropy $K_r$ being the same in the two cases. For complete resetting, we calculated both time-dependent moments and a perturbative steady state. For orientational resetting, we calculated the effective (anisotropic) diffusion tensor, which determines the effective Gaussian behaviour at late times.

\noindent In future works, it would  be interesting to study the first passage problem for an anisotropic diffusive particle in a two-dimensions in the presence of resetting. It would also be interesting to study resetting of other asymmetric particles, or to study the dynamics of resetting colloids in the presence of interactions. Intriguing effects of resetting could also be explored in three dimensions, both for rod-like particle with one orientational degree of freedom and for full anisotropic particles \cite{wittkowski2012self}.

\acknowledgements
\noindent SC acknowledges support from the Alexander Von Humboldt foundation. KSO and HL acknowledge support by the Deutsche Forschungsgemeinschaft (DFG) within the project LO 418/29-1.


%

\end{document}